\documentclass[aps,physrev,reprint,groupedaddress]{revtex4-2}

\usepackage[english]{babel}
\usepackage{graphicx}
\usepackage{xcolor}
\usepackage{dcolumn}
\usepackage{hyperref}
\usepackage{animate} 

\pdfoutput=1 
             
\def\be{\begin{equation}}
\def\ee{\end{equation}}
\def\bea{\begin{eqnarray}}
\def\eea{\end{eqnarray}}

\def\yzero{\smash{\hbox{$y\kern-4pt\raise1pt\hbox{${}^\circ$}$}}}

\def\beq{\begin{equation}}
\def\eeq{\end{equation}}
\def\beqa{\begin{eqnarray}}
\def\eeqa{\end{eqnarray}}

\def\-{\hphantom{-}}

\def\s2{\frac{1}{\sqrt2}}

\def\beq{\begin{equation}}
\def\eeq{\end{equation}}

\def\IF{\relax{\rm I\kern-.18em F}}
\def\II{\relax{\rm I\kern-.18em I}}
\def\IP{\relax{\rm I\kern-.18em P}}
\def\IC{\relax\hbox{\kern.25em$\inbar\kern-.3em{\rm C}$}}
\def\IR{\relax{\rm I\kern-.18em R}}

\def\Dsl{\,\raise.15ex\hbox{/}\mkern-13.5mu D} 
\def\IZ{Z\kern-.4em  Z}

\DeclareUnicodeCharacter{0301}{HERE}

\begin{document}

\preprint{APS/123-QED}

\title{\boldmath{Formative experience for intensive instruction physics courses: Evaluation and results in an Electromagnetism course}}

\author{Marcela Vallejo}
\email{mvallejo@ucn.cl}
\affiliation{Departamento de Física, Universidad Católica del Norte, 124000 Antofagasta, Chile}

\author{Ema Huerta}
\email{ehuerta02@ucn.cl}
\affiliation{UIDIN, Universidad Católica del Norte, 124000 Antofagasta, Chile}

\author{Joselen M. Peña}
\email{joselen.pena@uantof.cl}
\affiliation{Departamento de Física, Universidad de Antofagasta, 124000 Antofagasta, Chile}

\author{José Leiva}
\email{joseleivagu@santotomas.cl}
\affiliation{Escuela de Psicología,Universidad Santo Tomás de Antofagasta, 124000 Antofagasta, Chile}

\author{Ethan Rodriguez}
\email{ethan.rodriguez@alumnos.ucn.cl}
\affiliation{Licenciatura en Física, Departamento de Física, Universidad Católica del Norte, 124000 Antofagasta, Chile}

\date{\today}

\begin{abstract}

The rising demand for higher education has led universities to offer courses in multiple formats, including Intensive Instruction Courses (IICs), to meet the needs of a diverse student body. While active teaching methods improve physics understanding in standard courses, little research has examined their effectiveness in IICs. This research explored the most efficient methodologies for promoting meaningful learning in intensive physics courses. To this end, an integrated pedagogical proposal was designed based on the opinions gathered from a focus group of teachers with previous experience teaching these courses, as well as on existing literature, highlighting methodologies, types of assessment, and characteristics of ICCs. To evaluate its efficiency, a quasi-experiment was conducted in which students were divided into two groups: an experimental group (EG), which followed the teaching proposal, and a control group (CG), which received non-innovated lessons. Results were measured using the BEMA electromagnetism concepts inventory before and after the intervention. Statistical analysis revealed that Hake's Gain was greater in the EG, both in general terms and in each thematic unit of the course. The intervention in the EG showed that the use of active methodologies was more efficient than those used in the CG in the context of IICs. The results obtained suggest the need to continue investigating other factors involved in the teaching-learning process of IICs.
\end{abstract}

\maketitle


\maketitle
\flushbottom


\section{Introduction}  


In last decades, the university education sector has undergone significant changes, one of the most notable being the increase in demand for higher education, which has led to an increase in both the number and diversity of students, according to Schendel and McCowan in \cite{Schendel2016}. In Australia, for example, the government reported that between 1989 and 2014, the number of university students tripled \cite{Harvey2017}. This trend is reflected worldwide, and according to ICODE in \cite{i2015online}, it is projected that between 2000 and 2030, the student population will have quadrupled. Similarly, in 2024, UNESCO in its report \cite{UNESCO2024} states that the gross enrollment rate in higher education has increased globally, from 30\% in 2010 to 43\% in 2023, with particularly rapid growth in Latin America.


This increase in university enrollment, together with the needs of an increasingly diverse student body seeking to balance their studies with work and/or family commitments, has led universities to design study plans that are better suited to students' demands \cite{Burton2008}. One such approach is the intensive format \cite{Davies2006}, also known as Intensive Instruction Courses (IICs). These courses are developed in a shorter period of time than traditional courses, with fewer class hours, using different modalities, the first according to the form of interaction: synchronous, asynchoronous or hybrid; and the second according to the teaching environment: face-to-face, online or virtual, distance learning or combined.


In Latin America, various training programs have been developed through the IICs, most of which focus on language learning. However, intensive courses are also offered in other areas of knowledge, regulated in universities in countries such as Argentina, Bolivia, Venezuela, and Chile \cite{McGinn2020, UniversidadCatlicadeTemuco2021, ucn2024, Morales2021, ucab2025}. In Venezuelan universities \cite{UniversidadCentraldeVenezuela2005}, for example, intensive summer courses are designed to allow students to advance their studies or to retake failed subjects. They last between five and eight weeks, ensuring that the number of hours is equivalent to that of a regular semester. Similar examples can also be found in Chile, with intensive remedial courses in critical subjects (high-failure rate courses) offered at the Catholic University of the North \cite{ucn2024} and the Catholic University of Temuco \cite{UniversidadCatlicadeTemuco2021}, characterized by a duration of 3 weeks and exclusive dedication, i.e. students must devote their time to only one subject and cannot enroll in others at the same time.

\subsection{Background of Intensive Instruction Courses}


There are various definitions and applications for IICs, using terms such as "accelerated", "shortened", "blended", "flexible delivery", "block format" and "compressed" to describe them \cite{Davies2006}. According to Harvey et al. in \cite{Harvey2017}, IICs are defined as the delivery of a complete course in a shorter period of time than a traditional semester, while maintaining equivalence in learning outcomes and workload. In addition, these courses have different purposes, such as facilitating or accelerating students' academic progress or allowing non-traditional university students to balance family, work and studies through asynchronous online courses.


Regarding the quality-related attributes of IICs, Scott in \cite{Scott2003} mentions that these courses can reduce procrastination and	improve	academic preparation of students. Another element to consider is classroom interaction and	personalized attention from instructors, which is possible thanks to classes with fewer students than in a regular semester. In addition, he highlights that students tend to prefer active teaching	methods	and	memorable learning experience, and that it is necessary to emphasize key concepts and eliminate non-essential material to facilitate curricular continuity. Finally, it points out that the intensive nature of the course requires both students and teachers to maintain a steady pace of work, where timely	feedback is	a key factor in	 ensuring comprehension of the content in a short period of time.


In \cite{Daniel2000ARO}, Daniel points out that IICs tends to offer short-and long-term learning outcomes that are equivalent to, or even superior to, those of traditional courses in various disciplines. These better results could be attributed to the higher levels of motivation achieved by the students compared to traditional courses. Regarding the perceptions of students of this teaching modality, Karaksha et al. \cite{KARAKSHA2013BEN} suggest that IICs increase student satisfaction, with an increase observed as students progress in this modality, reporting positive experiences of the course and showing greater motivation and confidence in their learning \cite{Lee2010}. 


On the other hand, Manalo et al. \cite{Manalo1996} highlight the benefits of a four-day IIC at the beginning of the university term, with the aim of teaching study techniques such as effective time management, study organization, preparation for assessments, concentration, memory and written expression skills. Finally, Goode et al. \cite{Goode2023} recommend approaching IICs with a careful design, using active learning pedagogy, paying special attention to workload, a well-aligned curriculum and technology support, as well as offering opportunities for respectful and open dialog interactions. 

\subsection{Active Methodological Strategies in Higher Education for STEM Degrees} \label{estrategiasStem}


Higher education institutions have promoted changes in their pedagogical approaches, evolving from cognitive- based educational models towards socio-constructivist approaches that emphasise interaction and the collective construction of knowledge \cite{Freeman2014, Ghislandi_2014, Ramsden2003}. In this process, they have incorporated various methodological strategies to promote meaningful learning, with a special emphasis on science, technology, engineering, and mathematics (STEM) programmes, where active learning has been promoted as a way to strengthen the development of knowledge, skills, and abilities. According to Labrador and Andreu \cite{labrador2008metodologias}, active methodologies are methods, techniques and strategies that teachers use to transform teaching into a dynamic process that encourages student participation and enhances learning. Fraser \cite{Fraser2014} complements this view by pointing out that learning is considered active when students take a leading role in their own learning process.


Student-centred active methodologies have not only been shown to improve academic performance, but also contribute to reducing learning gaps among students from minority groups by increasing their self-efficacy in science \cite{Ballen2017}. Along these lines, other studies \cite{McDaniel2016, Schiltz2019} indicate that it generates significant improvements in conceptual understanding, although these do not always translate into a proportional development of problem- solving skills. Therefore, further research into its impact and effectiveness in different educational contexts is recommended.


There are various methodologies that promote meaningful and deep learning, some of which are described below:

\begin{itemize}


\item {\em Direct Instruction} \cite{Oakley2021}, is a methodology that uses three moments in the classroom: "I do it, we do it, you do it", progressively guiding the student towards autonomy in learning.


\item {\em Cooperative Learning} \cite{Johnson1999}, promotes cooperation among students to achieve common goals through interaction in small groups. For this methodology to be effective, it is essential that each member assume clear responsibilities and that active interaction be encouraged.


\item {\em Flipped Classroom}, is another methodology that has proven effective, allowing students to learn independently outside the classroom using resources such as videos and readings and then apply that knowledge in class. This approach, originally developed by Bergmann and Sams \cite{Bergmann2012}, has been proven to be effective in terms of learning achievement, motivation, self-efficacy, cooperation, and commitment \cite{HectorG2021}.


\item {\em Problem-Based Learning}, focuses on solving open-ended and complex problems in small groups \cite{Barrows1980}. This methodology promotes not only the acquisition of knowledge, but also the development of skills such as communication, teamwork and critical thinking, as well as enhancing problem-solving abilities.


\item {\em Formative Assessment}, consists of collecting and using evidence about student performance to inform both students and teachers about the level of learning achieved during a training session \cite{Black2009}, and plays a crucial role in the teaching-learning process. Formative assessment allows students to reflect on their own progress and teachers to adjust their teaching practices in a more informed manner. According to Bozzi \cite{Bozzi2021}, incorporating this strategy is highly effective in improving performance, especially among students with difficulties, as it allows the identification of whether effective learning is being achieved throughout the course \cite{tapia2023evaluacion}.

\end{itemize}

\subsection{Methodologies for Teaching Physics} \label{estrategiasFisica}


Despite evidence supporting the effectiveness of active methodologies, there is still considerable resistance among teachers to its implementation, due to several factors, such as: they require more time and resources \cite{deslauriers2019measuring, silverthorn2006s}, more training for teaching staff, and a reduced number of students in the classroom. Therefore, the most common approach is to opt for traditional lectures \cite{stains2018anatomy,henderson2007barriers}, which are part of a teacher-centered methodological approach, where the needs and motivations of the teaching staff are prioritized over those of the students, based predominantly on traditional lecture teaching, with little student participation and limiting the development of key skills such as reflection, creativity, communication and collaborative work.


Currently, learning science, and physics in particular, represents a major challenge for many students in STEM faculties \cite{Angell2004, ornek2007makes, redish2006reverse}. In the context of active methodologies developed for teaching and learning physics, there are various techniques and strategies in the literature, among which the following stand out: the {\em Peer Instruction methodology} \cite{Crouch2001}, the {\em Tutorials in Introductory Physics methodology} \cite{McDermott1999}, the {\em use of Simulations} \cite{guaman2023uso}, {\em Problem-Based Learning} \cite{duch1996problems}, and {\em Collaborative Work} \cite{Heller1992, Springer1999, Stamovlasis2006}. These methodologies have been developed mainly in North American universities and yet, despite the consensus on their effectiveness, they are not widely used in physics courses \cite{stains2018anatomy}. Most of the active methodologies mentioned above share the characteristic of being flexible and can complement or replace traditional classes, with the teacher guiding and supervising discussions to ensure understanding before moving on to other topics.

The methodology known as {\em Peer Instruction}, introduced by Mazur in 1991 \cite{Mazur:0}, has proven to be very efficient. In this methodology, the teacher first distributes the material related to the class, then asks conceptual questions, which must be answered individually. Based on the percentage of correct answers, the teacher then defines the actions to be taken.


As for the strategy of {\em Tutorials in Introductory Physics}, it is described as consisting of three complementary activities: a pretest, the application of the tutorial, and additional exercises. Benegas in \cite{benegas2007tutoriales} reports the use of this teaching-learning strategy in first-year university courses, with the main objective of developing conceptual understanding of physics.


The use of {\em Online Simulation} tools, for example \cite{ophysics2025}, is another effective tool for teaching physics, allowing students to explore and understand abstract physical concepts interactively. According to Guaman's research \cite{guaman2023uso}, online simulators are beneficial for improving and stimulating learning, but they should not completely replace the traditional laboratory experience. Their implementation must be aligned with learning objectives, and teachers must plan their pedagogical use to maximise their effectiveness in the classroom.

\subsection{Definition of the problem and motivation}


IICs are often questioned for allegedly lowering the quality of learning and academic rigor during their training activities. Nevertheless, many institutions continue to offer these courses, whose methodological and curricular design is based more on assumptions and traditions than on solid empirical evidence \cite{Scott2003}. In this vein, Ontong \cite{Ontong2022} has presented research results suggesting that there are no statistically significant differences between the grades of students who repeat a subject in an IIC compared to a regular course. This indicates that intensive or accelerated learning can be an effective way to help students who have failed a subject to accomplish the expected learning outcomes. However, further research is required to identify the most appropriate pedagogy for science teaching and its adaptation to the undergraduate  IICs format, based on empirical evidence, as mentioned by Harvey et al. \cite{Harvey2017}, who emphasize the need to expand research in this area. This raises the need to investigate the effective integration of active methodologies for teaching physics in IICs at the undergraduate level, as well as to ensure that students not only knowledge acquisition but also develop critical and lasting skills in a constantly changing educational environment.


In order to advance the understanding and enrichment of IICs, this study aims to contribute to the search for an efficient teaching approach for intensive physics instruction. The questions that guided the research were as follows:

\begin{itemize}

\item Q1.- What are the most appropriate characteristics, assessment strategies, and teaching methodologies for development of a physics IIC in STEM degree programs?


\item Q2.- Are there differences in student performance depending on the teaching strategies designed for different groups (experimental group and control group)?


\item Q3.- Is the innovative teaching approach based on the use of active methodological strategies and modern assessment more efficient for learning electromagnetism in a IIC?

\end{itemize}


To answer these research questions, this paper is organized as follows: Section II describes the stages of the research and the procedures and materials used in each stage, particularly in the development of the quasi-experiment. Section III reports the results obtained in both the qualitative and quantitative phases, as well as a description of the statistics applied. Section IV discusses the results. Finally, Section V presents the conclusions.

\section{METHODS}


Taking into account the above considerations, the objective of this research was to evaluate the teaching efficiency of an innovative training proposal for an Intensive Instruction Course (IIC) on Electromagnetism, through the integration of active methodological strategies and modern assessment, using a descriptive approach and mixed research methodologies that included the development of a focus group, literature review, and the development of a quasi-experiment.


The research was conducted at a Chilean university and was approved by the scientific ethics committee of the same institution through resolution 086b/2023. IICs at this university are offered during the winter and summer terms, focus on critical subjects (high-failure rate courses) and are considered "remedial courses", meaning that this modality allows students who fail a course in a regular semester to re-enroll in the subject in an accelerated format instead of repeating it over a regular semester. IICs last three weeks, for a total of 45 hours, including three hours of direct teaching per day, divided into two blocks with a break between them. This corresponds to approximately 40\% less time than a regular course and requires exclusive dedication on the part of the student.


The participation of teachers and students in the research was voluntary, and they signed an informed consent form in accordance with the institution's ethical guidelines. Only one student in the control group decided not to participate and was therefore excluded from the analysis. The research was implemented in three stages, which are described below:

\subsection{Stage 1}
%

In this first stage of the research, the most effective methodological strategies for an IIC in the STEM area, as well as for teaching physics, were identified through an exhaustive review of the literature and the development of a focus group with teachers who have previously taught this type of course.

%
%

\begin{figure}
\begin{center}
\begin{tabular}{ c }
 \includegraphics[width=0.45\textwidth]{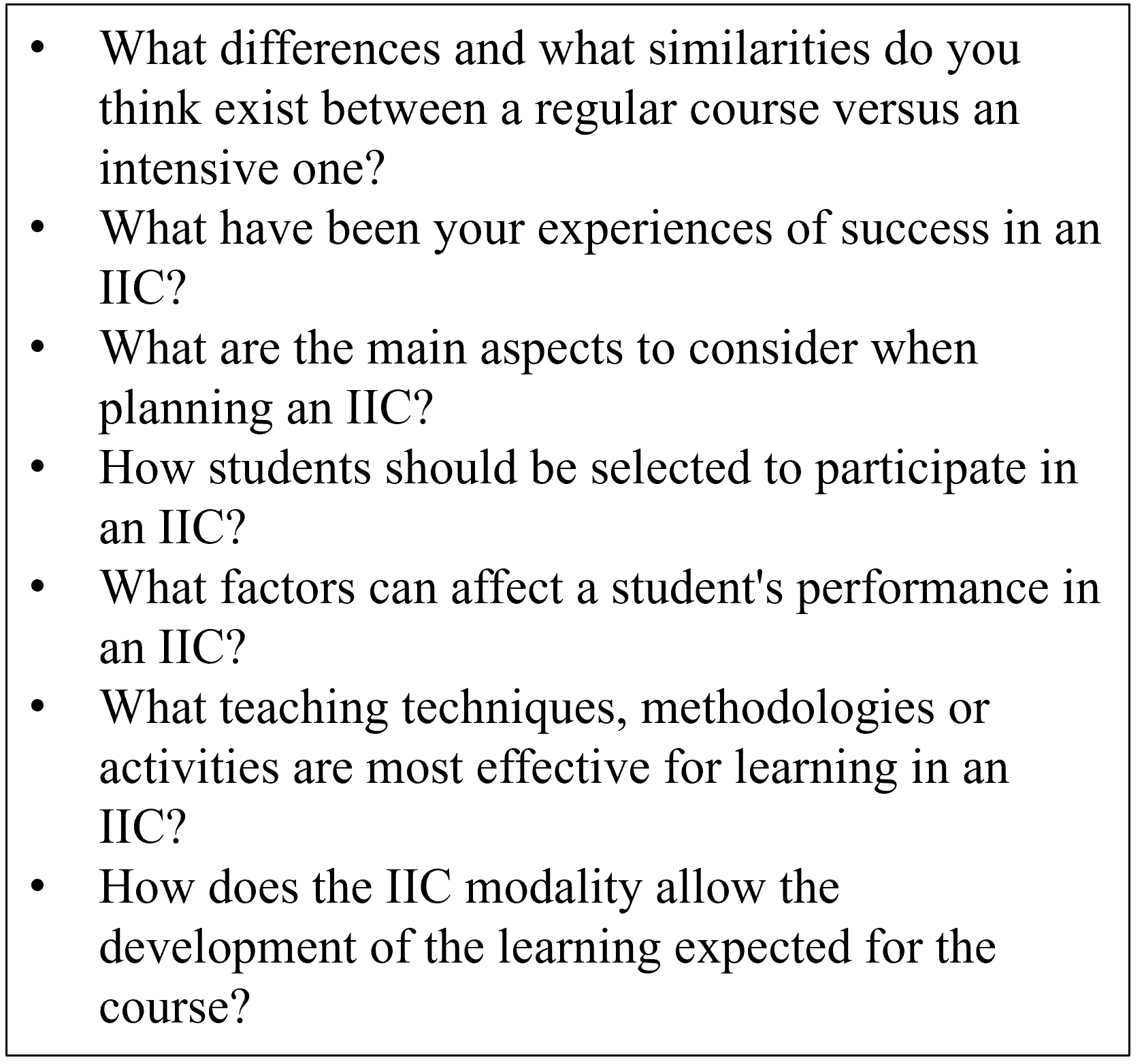} \\     
\end{tabular}
\end{center}
     \caption{This table presents the questions that guided the focus group development.}
    \label{cuadro1}
\end{figure}

The objective of the focus group was to identify the characteristics, teaching strategies and methodologies most recommended by teachers with previous experience teaching IICs at the institution. To this end, carefully designed open-ended questions were formulated, in order to guide the discussion and explore in depth the opinions of the teachers. The thematic script that guided the focus group discussion is shown in Figure \ref{cuadro1}.

The focus group included six university teachers with master's and doctoral degrees in their respective disciplines, as well as training in university teaching through courses, workshops and diploma programs. All of them had taught IICs in  the areas of Mathematics and Physics at the institution for various STEM degrees on at least two occasions, in addition to having previous experience in university teaching ranging from 2 to 33 years. Each participant met the inclusion criteria related to previous experience in IICs, teacher training, and teaching practice in higher education. The activity was conducted virtually via the Zoom platform and was recorded with the consent of the participants. An invitation was sent to the teachers to organize the activity, which was led by a moderator with the support of two observers and lasted 120 minutes.


For the analysis of the information collected in the first instance, the entire conversation was transcribed from audio to text. The speeches of the participants were analyzed and organized according to the proposal by Glaser and Strauss \cite{glaser1967discovery}, identifying the main areas according to the thematic script.

\subsection{Stage 2}


In the second stage of the research, a training proposal was developed for STEM students for a IIC on electromagnetism, based on both the information collected through the focus group and an exhaustive review of the literature. The pedagogical proposal included some of the active methodological strategies described in sections \ref{estrategiasStem} and \ref{estrategiasFisica}, among which the following were selected: interactive lectures, direct teaching, collaborative work, problem-based learning, use of online simulation tools, and formative assessment. In the appendix, we can find, as an example, the log of activities based on the teaching plan for the 2024 intensive summer course on electromagnetism for the course where the intervention is carried out (Experimental Group, EG).


In addition to developing the pedagogy proposal, at this stage the assessment tool to be used during the quasi- experiment was selected, as well as the indicator to quantify and comparatively evaluate the level of learning achieved by the students. The assessment instrument selected was the BEMA (Brief	Electromagnetism Assessment) test, which was developed by Ruth Chabay and Bruce Sherwood \cite{ding2006evaluating} and is a {\em Concept Inventory (CI)} widely used in Physics Education Research (PER). This test is a standardized, validated and reliable instrument for assessing students' qualitative understanding of key concepts in electricity and magnetism, covering the topics of electrostatics, direct current circuits, magnetostatics and electromagnetic induction. It consists of 31 multiple-choice questions with variable options, between three and ten alternative answers. The reliability of the BEMA has been examined in multiple studies, by various authors, and in various countries \cite{xiao2019linking, hansen2021multidimensional, eaton2019comparing, ding2006evaluating}.


To quantify and evaluate the level of learning achieved by students, the coefficient known as {\em Average Normalized Gain(g)} or {\em Hake's Gain} was selected, defined by Hake in \cite{Hake1998} as the ratio of the increase in the percentage of correct answers (percentage of achievement) between the pre-test $(\%CA_{Pre)}$ and the post-test $(\%CA_{Post})$ with respect to the maximum possible achievement percentage, see equation (\ref{Hake_Gain}). The use of (g) allows for the evaluation of progress, both individually for each student, thus avoiding comparisons between students with different initial levels of knowledge, and as a group, using average achievement percentages. Subsequently, Covian and Celemin \cite{CovinRegales2008} redefine (g) as the {\em Didactic Efficiency} of a course and relate it to the influence of the teaching-learning process on conceptual understanding, establishing it as an absolute value that varies in the range of 0-1, also being used in research as a percentage value. It should be noted that Hake's Gain for a course serves not only to evaluate the influence of the teaching-learning process, but also to assess the efficiency of a study program \cite{kohlmyer2009tale}.


For this research, we will use Hake's Gain, also considering the interpretation proposed by Covian and Celemin \cite{CovinRegales2008}, given the relationship between these indicators and the objective of the study.


In the literature, there are two methods for calculating Hake's Gain \cite{Hake1998}, and it is not always explicitly stated which one is used. The first approach, defined by Hake, calculates the gain (g) using the average of the achievement percentages between the pre-test and post-test 
defined by equation:
\begin{equation}\label{Hake_Gain}
g = \frac{<(\%CA_{Post})> -<(\%CA_{Pre}) >}{100-<(\%CA_{Post})>}
\end{equation}
with initial percentage of correct answers $(\% CA_{Pre})$ and final percentage of correct answers $(\% CA_{Post})$. Also, $<(\%CA_{Pre})>$ and $<(\%CA_{Post})>$ indicate the averages of each quantity. In addition, this concept can be used to calculate the {\em Learning Individual Gain} $(g_i)$ 

\begin{equation}\label{Individual_gain}
g_i = \frac{(\%CA_{Post})_i-(\%CA_{Pre})_i}{100-(\%CA_{Post})_i}
\end{equation}
with initial percentage of correct answers $(\%CA_{Pre})_i$ and final percentage of correct answers $(\%CA_{Post})_i$ for each student $i$. The second approach, on the other hand, uses {\em Mean of Individual Gains} ($<\text{g}>$), i.e, 
\begin{equation}\label{Mean_of_Individual_Gains} 
<g> = \frac{\Sigma_i g_i}{N}
\end{equation}
where N is the total number of students in the sample. According to Hake and Bao \cite{Bao2006}, the difference between the two calculations is not significant in classes with a large number of students, although there may be variations in small groups. In an analysis conducted by Von Korff et al., \cite{von2016secondary} it was found that most studies reporting Hake's Gain do not specify which of these two methods was used. Although Hake defined it using the first approach, many researchers resort to the second, and he also showed that the differences between the two approaches rarely exceed 5\%. For this reason, both methods are accepted and widely used in PER. It should be noted that although both Hake's Gain (g) and Didactic Efficiency were initially defined to assess the understanding of force concepts through the FCI force concept inventory test, in this research we extend their application to assess the concepts of electricity and magnetism through the BEMA test.

\subsection{Stage 3}

\begin{figure}
\begin{center}
\begin{tabular}{ c }
 \includegraphics[width=0.45\textwidth]{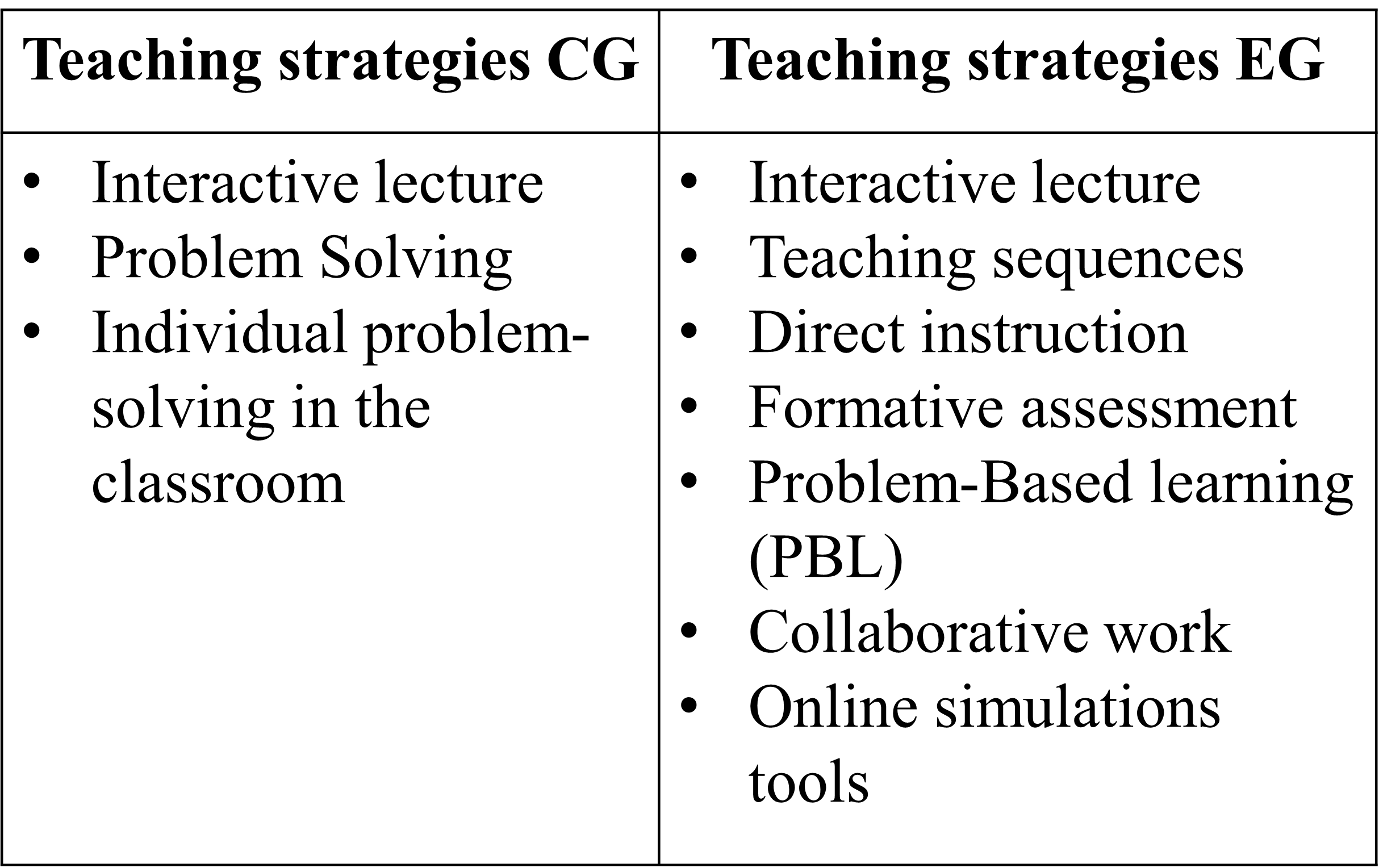} \\
\end{tabular}
\end{center}
     \caption{Comparative table of teaching strategies used for the control CG and experimental EG groups.}
    \label{tabla1}
\end{figure}

In the third stage of the research, a quasi-experiment was carried out to validate the innovative methodological proposal established during stage 2. To this end, two groups of students were formed: an experimental group (EG) where the proposed pedagogical innovations were applied, and a control group (CG), which followed a non-innovative methodology. Figure \ref{tabla1} presents a comparative overview of the teaching strategies used in the CG and the EG.

%


The participants in this stage were the 41 students enrolled in the Electromagnetism summer IIC, during January 2024, with 20 students in the GE and 21 students in the GC. The participating students could not be randomly assigned to the experimental group, as they were defined based on the order in which they enrolled in the course. The GC and GE classes were taught by two different teachers at the same time, with a schedule for the GE organized as follows: each day had two 90-minute class blocks, with a 20-minute break between them. Each 90-minute block was subdivided into two 40-minute sessions with a 10-minute break between them. For more details, see the appendix \ref{apéndice} that shows the log of activities carried out in the EG.


The BEMA test was administered at two points during the course: at the beginning of the course (pre-test) and at the end of the course (post-test), which were carried out in both the CG and EG groups. To identify whether the innovative methodology had an effect on performance in the BEMA test, the mean values of the percentages of correct answers in the pretest and post-test were compared, both between courses to identify whether there were differences at the beginning and end, and within courses to determine whether there were significant changes. To do this, the Shapiro-Wilk test \cite{shapiro1965analysis} was first used to identify normality. The Wilcoxon test \cite{wilcoxon1950some} was used for measurements within each group, and the Mann-Whitney U test \cite{Mann1947} was used for comparisons between groups, except in cases where normality was present, in which case the Student's T-test \cite{biometrika1908probable} was used for independent samples. In this way, we evaluated whether the initial and final measurements were different between the CG and the EG, as well as possible improvements over time.

\section{RESULTS} 


The results obtained in each phase of the project are presented below, beginning with the results of the focus group and then presenting the results obtained in the quasi-experiment.

\subsection{Qualitative Phase: Focus group results}


According to the results obtained in the focus group, five areas of the IICs were identified (Figure \ref{esquemagrupofocal}), the first of which related to their characteristics, the second to the most effective learning methods, the third to the form of assessment, the fourth to the characteristics of the teacher, and the fifth with the characteristics of the student body. These are described below:

\begin{itemize}


\item {\em Regarding the characteristics of IICs:} teachers perceive that they progress rapidly, covering the programme content in a period of three weeks, unlike a regular course, which takes a semester. Another characteristic mentioned is exclusive dedication, as each student must devote their time to only one subject. A third characteristic is the period in which these courses are taught, as they take place during the summer or winter break. Along the same lines, it is noteworthy that IICs have a smaller number of students compared to a regular course, with an average of 20 students, unlike a regular course, which has 50 students on average in institutional subjects. Another characteristic is that, as the course progresses, the content is addressed in a prioritized manner, which for the teaching staff means placing greater emphasis on topics that will be relevant for subsequent courses.


\item {\em Regarding the most effective learning methods}: teachers highlight collaborative work through problem solving and classroom exercises. They also consider guided exercises, carried out jointly by teachers and students, to be a relevant methodology. Finally, they mention the use of technology as a teaching support tool during classes through simulations and the One Note platform.


\item {\em Regarding the form of assessment}: they mention that there are different types of assessment implemented in the IICs, such as diagnostic, formative, and summative assessments. They also highlight that summative assessments are weekly, unlike regular courses, which are monthly.


\item {\em Regarding the characteristics of teachers in IICs}: it is essential that they have previous experience teaching the subject during a regular semester, as this allows them to acquire greater expertise on the content and the pace of student learning. Another important aspect is the more personalized attention with the students, which gives the teacher a better understanding of the level of knowledge and the pace of progress, both individually and as a group, in the learning process, given that the courses usually have a small number of participants. In addition, it is considered important for the teacher to regulate the progress of the course gradually, beginning with less difficult content and then moving on to more difficult topics.


\item {\em Regarding the characteristics of students participating in IICs}: first, it mentions that they show greater motivation to learn compared to a regular course, demonstrating more interest and participation in class. They also point out that they have prior knowledge of the topics that will be covered in the course, as they previously took the subject and failed it in the previous semester. Along the same lines, they point out that they have varying levels of knowledge about the subject, with some students having high levels and others low levels in the same course. Finally, another characteristic is that they decide to enroll in the course voluntarily and independently. 


These results are summarized in Figure \ref{esquemagrupofocal}.

\begin{figure}
\begin{center}
\begin{tabular}{ c }
 \includegraphics[width=0.47
\textwidth]{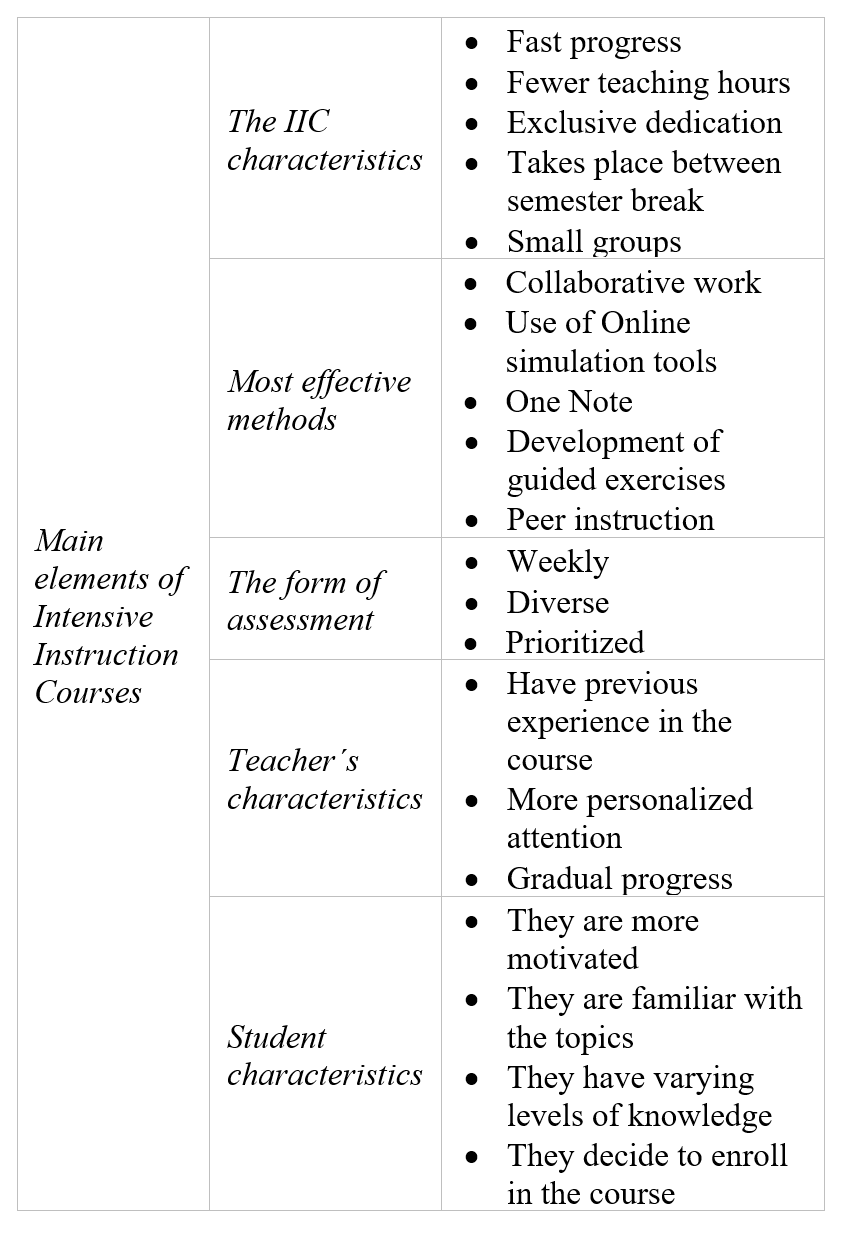} \\
\end{tabular}
\end{center} 
     \caption{Main elements of Intensive Instruction Courses IICs.}
    \label{esquemagrupofocal}
\end{figure}

\end{itemize}
%

\subsection{Quantitative phase. Results of the quasi-experiment}


This section describes the results of the statistical study of the research, which compares two groups: a control group (CG) and an experimental group (EG) that underwent the intervention. Descriptive statistics and normality tests were initially applied, followed by group comparison analyzes.

\subsubsection{Descriptive Statistics and Normality Tests}
 
\begin{figure}
\begin{center}
\begin{tabular}{ c }
\includegraphics[width=0.46\textwidth]{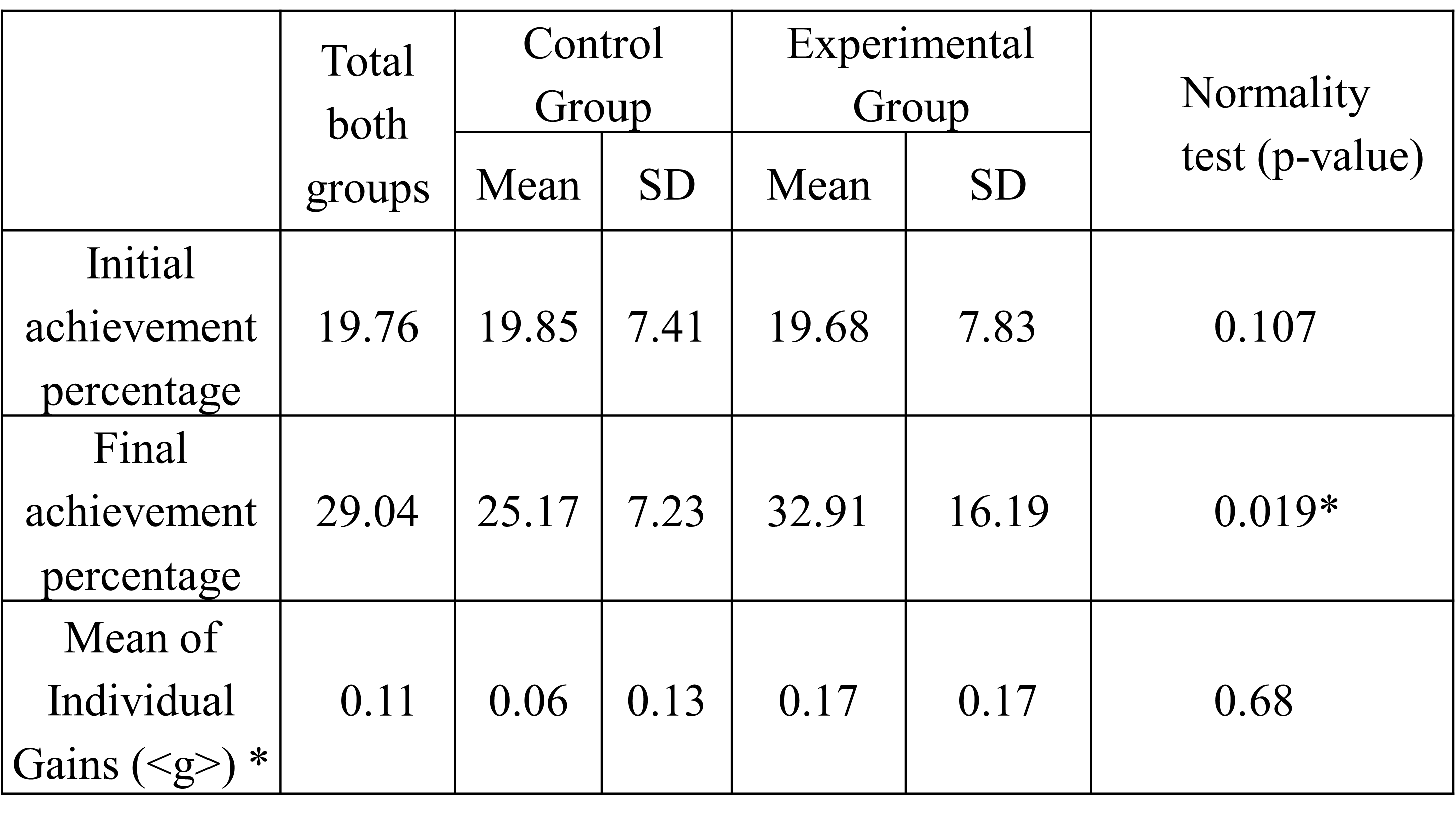} \\     
\end{tabular}
\end{center}
    \caption{ Statistical indicators for the experimental and control groups. *Note that the p-value for the post-test achievement percentage is less than 0.05.}
  \label{tabla2}
\end{figure}
\begin{figure}
\begin{center}
\begin{tabular}{ c } \includegraphics[width=0.4\textwidth]{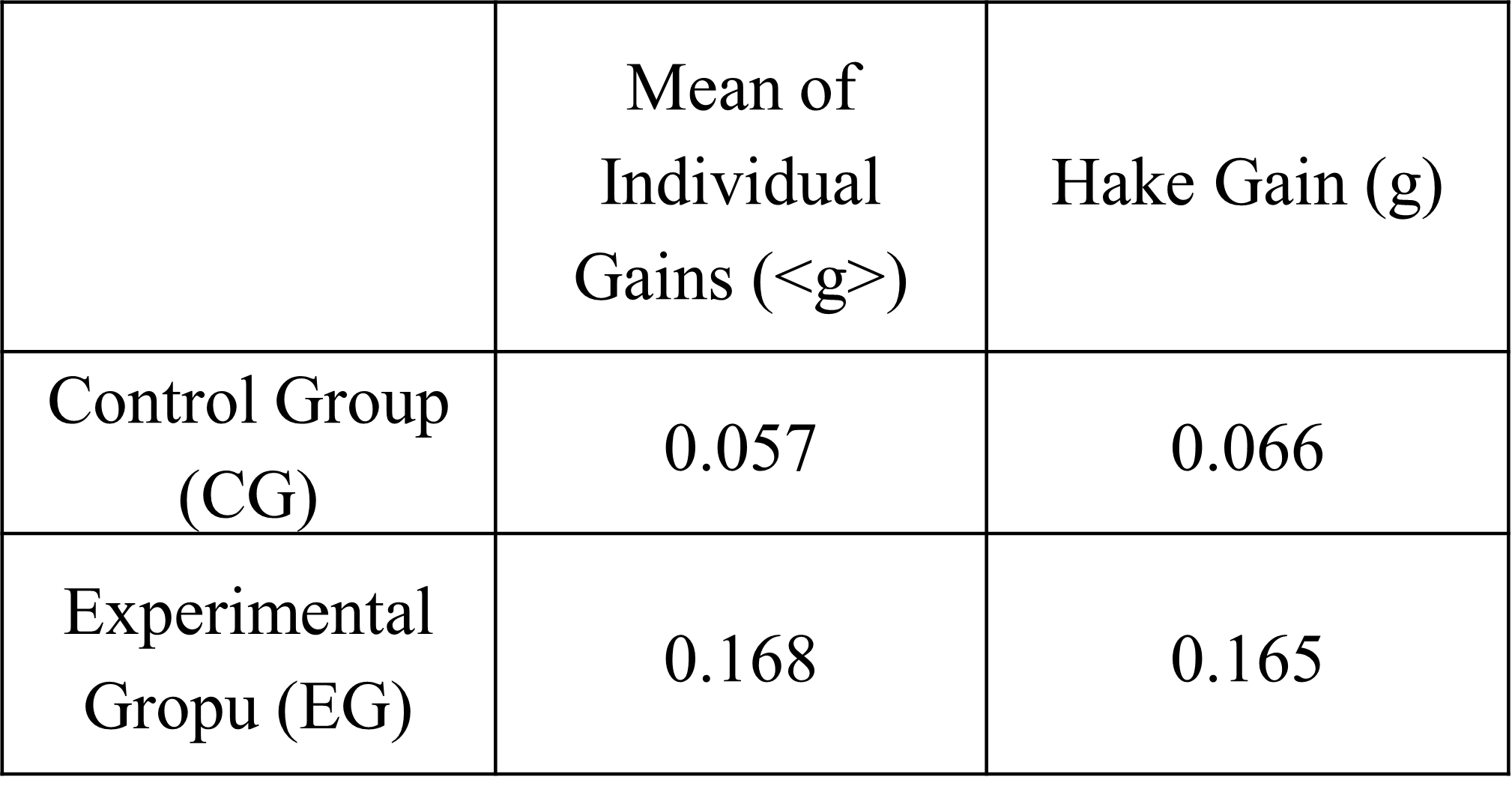} \\     
\end{tabular}
\end{center}
     \caption{Comparison of Mean of Individual Gains and Hake's Gain by group.}
    \label{tabla3}
\end{figure}

The means and standard deviations (SD) of the achievement percentages corresponding to an initial assessment (pre-test) and a final assessment (post-test) were calculated for both groups. The results are presented in Figure \ref{tabla2}. In addition, Mean of Individual Gains (Equation \ref{Mean_of_Individual_Gains}) ($<\text{g}>$) was determined and allow testing the mean differences. These results are reported in the same table. Furthermore, the Hake's Gain (g), which provides an indicator at the course level and whose interpretation is strictly descriptive, is shown in Figure \ref{tabla3}. It is important to note that both (g) and ($<\text{g}>$) are used to measure the efficiency of a course in conceptual understanding; see Bao in \cite{Bao2006}, and in general, these two calculations yield very similar results.

Figure \ref{tabla2} shows that the mean percentages of achievement  in the initial assessments $<(\%CA_{Pre})>$ for CG and EG are very similar. However, in the latter group, the mean post-test achievement percentage $<(\%CA_{Post})>$ is higher. In the case of the Mean of Individual Gains ($<\text{g}>$), this is higher in the EG compared to the CG. Regarding the normality test, only the post-test achievement percentage was not distributed according to a normal curve (p=0.019). 


Figure \ref{tabla3} shows Hake's Gain (g) for each group and compares it with the Mean of Individual Gains ($<\text{g}>$), showing that both are very similar in the EG, differing more in the CG. In the case of the Hake's Gain (g) indicator for each group, this can only be studied descriptively, since it is not possible to perform group difference tests with this calculation. It should be noted that this indicator is also higher in the EG, which implies that, in terms of overall performance gains from the course, it is superior to the CG. In other words, it can be said that the EG had greater gains in both indicators (g) and ($<\text{g}>$).

\subsubsection{Comparison between the control group (CG) and the experimental group (EG)}


The initial and final achievements of each group were compared. Considering the non-normality of the post-achievement percentage, the Mann Whitney U test was used. The Student's T test was used for the initial achievement.

\begin{figure}
\begin{center}
\begin{tabular}{ c }
\includegraphics[width=0.4\textwidth]{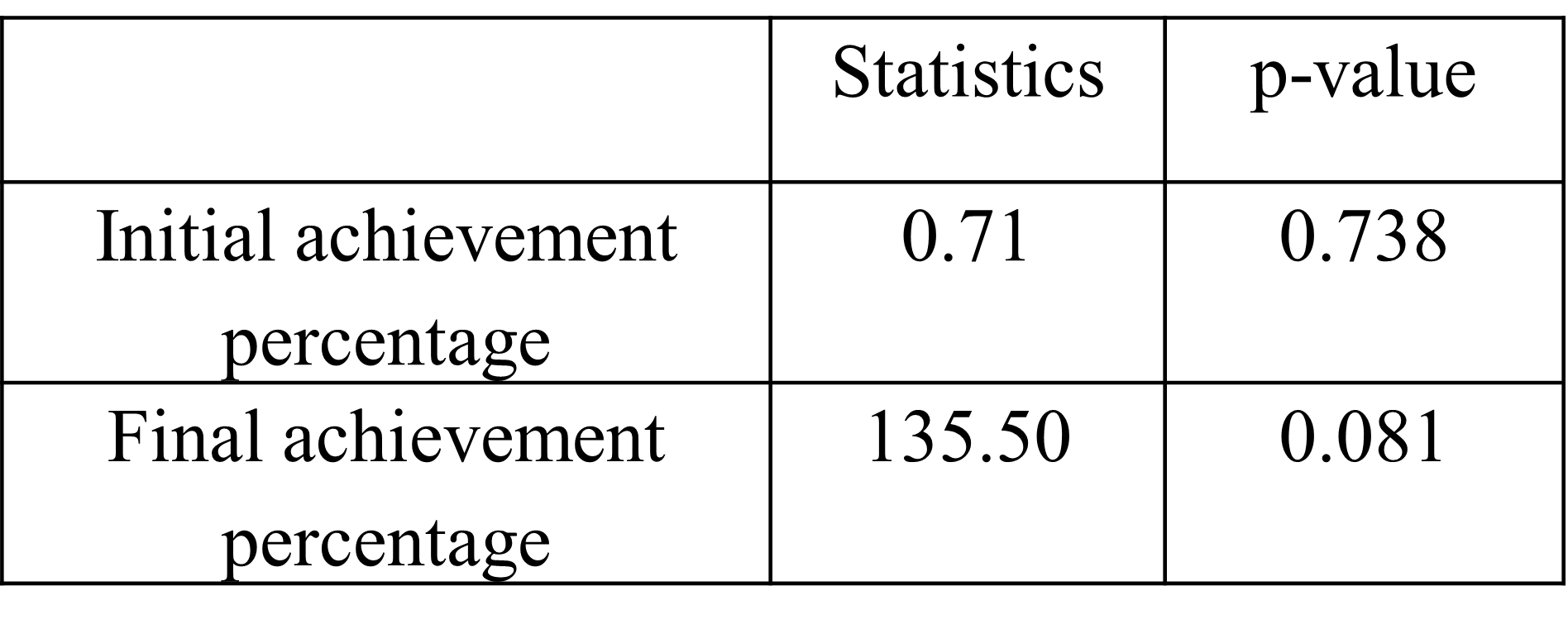} \\     
\end{tabular}
\end{center}
     \caption{Statistical indicators used to compare pre- and post-test achievement percentages in the groups.}
    \label{tabla4}
\end{figure}

Figure \ref{tabla4} shows that no significant differences were found in the achievement percentages when comparing CG and EG. In this sense, the levels initially obtained by each group are not different, nor are the results at the end of the course. However, a much lower p-value is observed in the post-test and a much more pronounced difference.

\subsubsection{Comparison within each group}

\begin{figure}
\begin{center}
\begin{tabular}{ c } \includegraphics[width=0.4\textwidth]{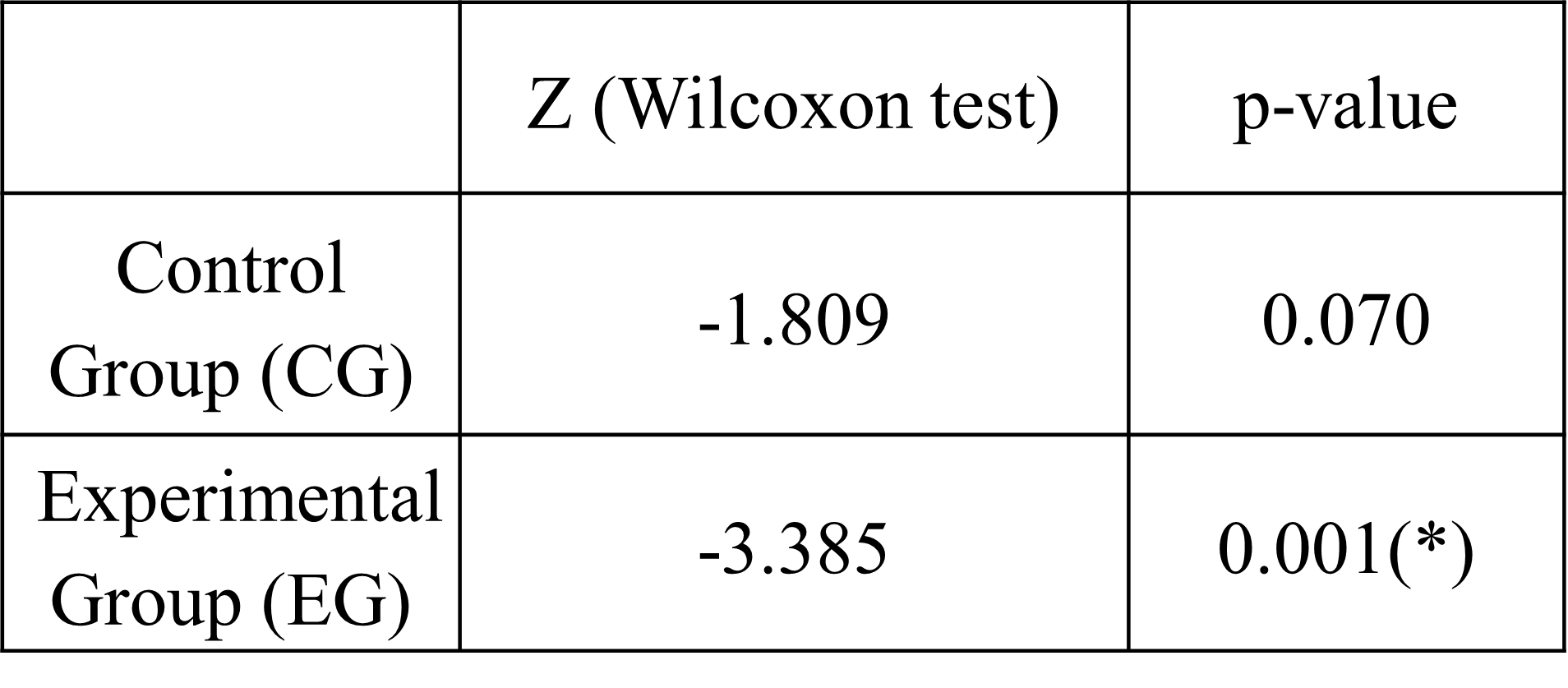} \\     
\end{tabular}
\end{center} 
     \caption{Wilcoxon test for experimental and control groups. Comparison of pre-and post-test achievement percentages for each group.}
    \label{tabla5}
\end{figure}

Given that comparisons over time within each group involve a non-parametric measurement, the Wilcoxon test was used. These analyzes are presented in Figure \ref{tabla5} and Figure \ref{tabla6}.
%


Figure \ref{tabla5} shows that in the CG, no significant differences were found between the initial and final percentages of achievement.This means that its performance has changed little over time. However, in EG, there was a significant difference ($p=0,001$), which implies that the students had a higher percentage of achievement in the post-test than in the pre-test much higher in the end, i.e., their performance changed over time.


According to the results obtained, it can be stated that intervention with active methodological strategies and modern assessment in the GE proved to be more efficient in learning basic concepts of electromagnetism, answering research question Q3.
\newline

\subsubsection{Comparison of Mean of Individual Gains}


The Mean of Individual Gains between the two courses were compared using a parametric test, given that this variable was normally distributed.

\begin{figure}
\begin{center}
\begin{tabular}{ c }
\includegraphics[width=0.45\textwidth]{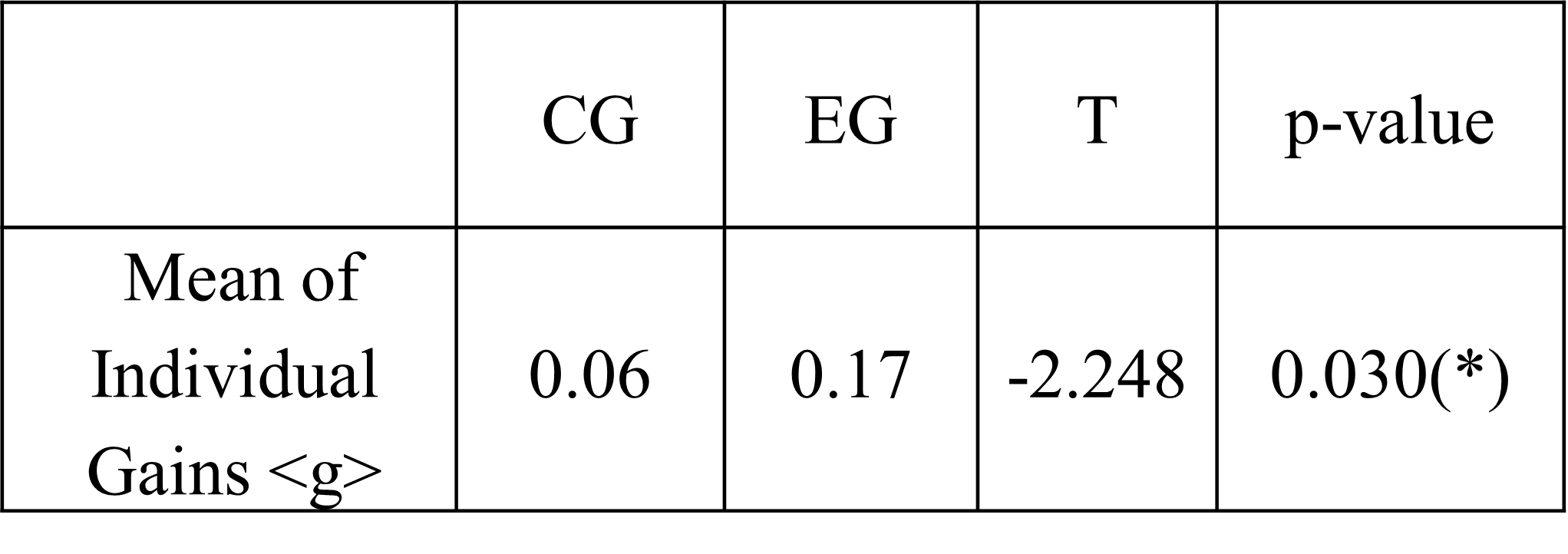} \\     
\end{tabular}
\end{center} 
     \caption{Comparison of Mean of Individual Gains ($<\text{g}>$) for the experimental group GE, control group GC, p-value, and T-test.}
    \label{tabla6}
\end{figure}

It was observed that the Mean of Individual Gains (Figure \ref{tabla6} ) was found to be significantly higher in EG (p=0.030), which implies that of the total that could have been improved, this group was able to improve more. Thus, it can be observed that although, in the end, the achievement percentages are not significantly different between the courses, the intervention group does have a significant improvement and an equally higher gain. This means that the average performance is better in the EG and that these are not attributable to chance. These results partially answer the research questions Q2 and Q3.

\subsubsection{Hake's Gain by thematic unit}

\begin{figure}
\begin{center}
\begin{tabular}{ c }
 \includegraphics[width=0.45\textwidth]{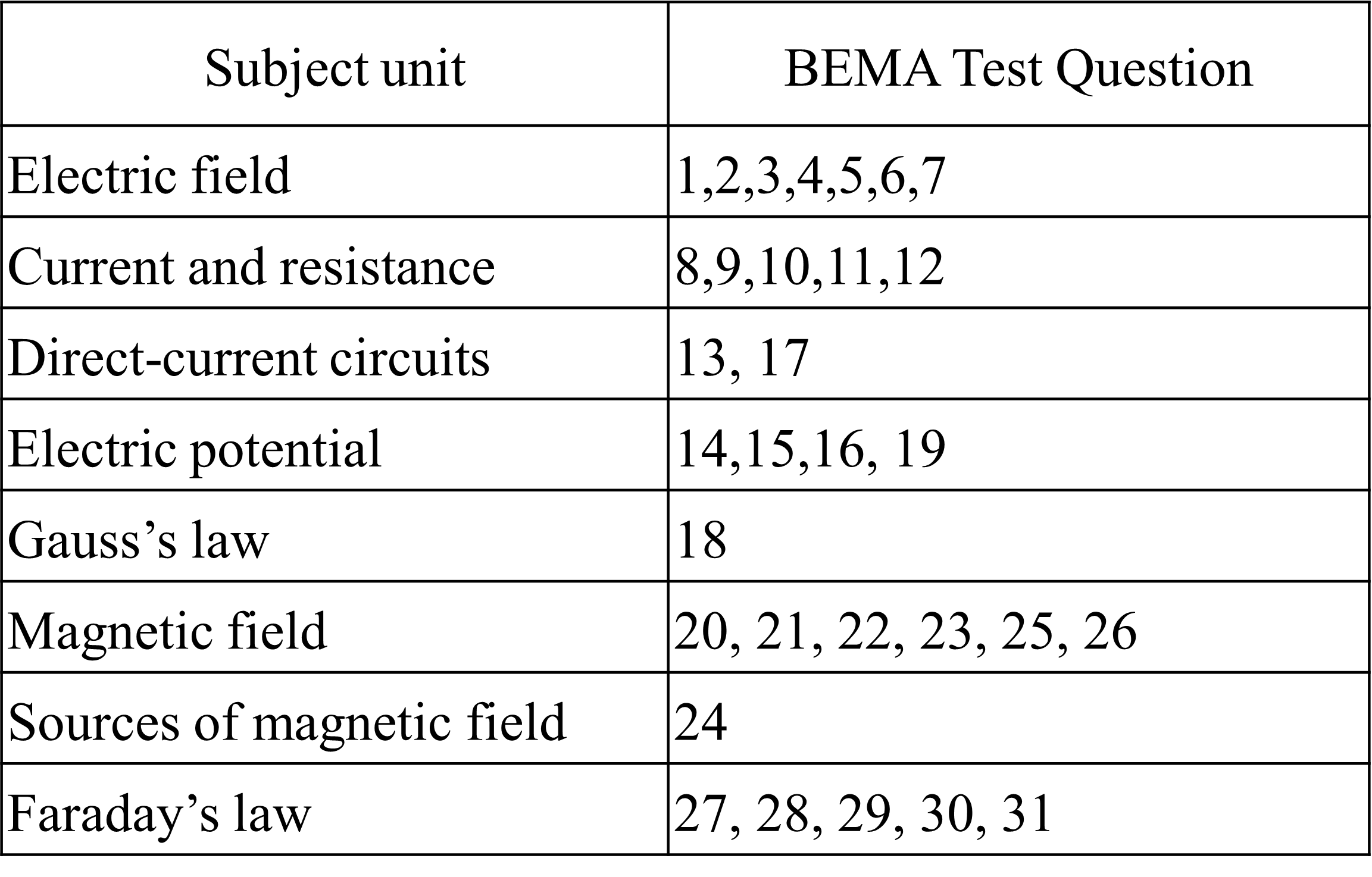} \\     
\end{tabular}
\end{center}
    \caption{Classification of BEMA test questions by subject units.}
     \label{tabla7}
\end{figure}

This section provides a comparative overview of the percentage of correct answers obtained in the different thematic units defined for the intensive Electromagnetism course in both EG and CG, classifying the questions from the BEMA concept inventory according to the thematic units of the course (Figure \ref{tabla7}).

\begin{figure}
\begin{center}
\begin{tabular}{ c } \includegraphics[width=0.5\textwidth]{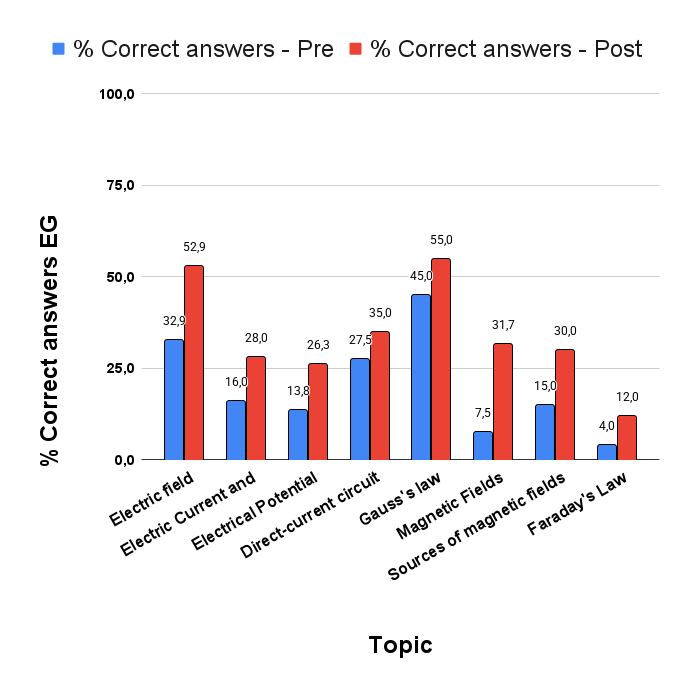} \\   
\end{tabular}
\end{center} 
     \caption{Percentage of correct answers per thematic unit in EG.}
     \label{grafico1}
\end{figure}

\begin{figure}
\begin{center}
\begin{tabular}{ c }
 \includegraphics[width=0.54\textwidth]{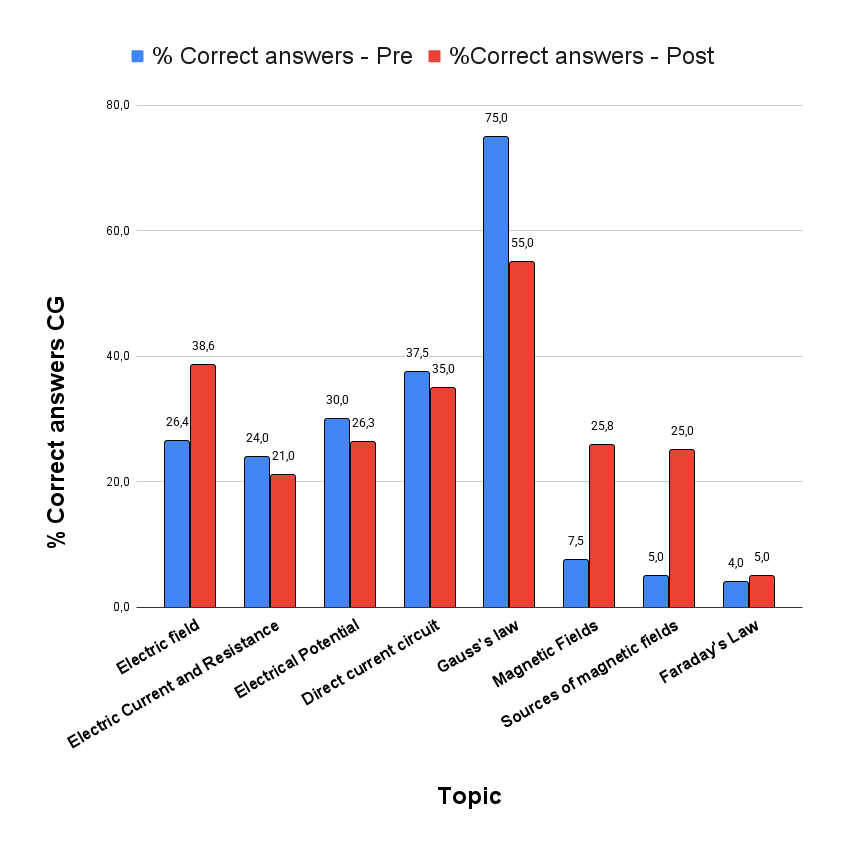} \\     
\end{tabular}
\end{center}
     \caption{Percentage of correct answers by thematic unit in CG.}
    \label{grafico2}
\end{figure}

Figures \ref{grafico1} and \ref{grafico2} show bar charts where we can find the percentage of achievement (or percentage of correct answers) per subject area in the pre-test and post-test for GE and GC respectively. Furthermore, these graphs show that in all thematic units in the EG, there is a greater relative increase in the percentage of correct answers compared to the CG.

\begin{figure}
\begin{center}
\begin{tabular}{ c }
 \includegraphics[width=0.5\textwidth]{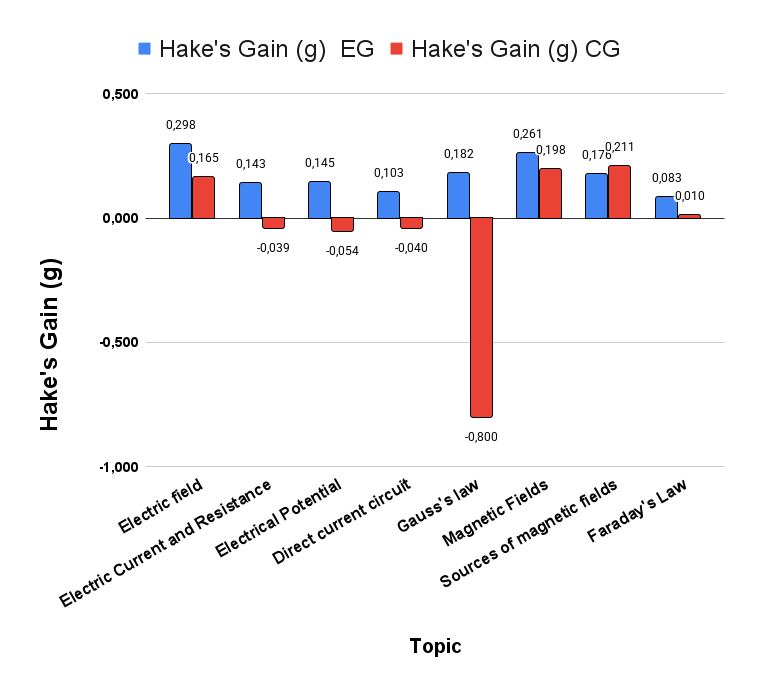} \\     
\end{tabular}
\end{center}
     \caption{Hake's Gain (g) by thematic units in the EG and CG.}
     \label{grafico3}
\end{figure}

In addition to the above indicators, it is also interesting to compare the Hake's Gain by thematic unit  between the two groups, which can be seen in Figure \ref{grafico3} . 

In Figure \ref{grafico3}, we can see that the Hake's Gain for all thematic units are higher for the EG. In addition, the Electric Field and Magnetic Field units show a significant Hake's Gain (close to 0.3) in learning \cite{ramirez2014aprendizaje}.

\twocolumngrid

\section{DISCUSSION}

%
When reviewing the results obtained in stage 1 through the focus group, it can be seen that the participants highlight the use of active methodologies for the development of learning in the IICs, which are characterized by being student-centered, encouraging greater involvement of students in their learning process, and promoting meaningful and in-depth learning. This is in line with Ballen's \cite{Ballen2017} approach, who also points out that these methodologies help reduce learning gaps, improve academic performance, and self-efficacy. In other words, these methodologies place students at the center of their own learning, promoting autonomy, motivation, and critical thinking, see Schiltz et al. \cite{Schiltz2019}, in line with current trends in higher education. Another element to consider is the use of technologies such as digital notebooks (OneNote) and online simulation of physical situations, which are effective tools that allow students to understand abstract physical concepts, generating greater interest and motivation for learning among students, in line with what Rosales et al. propose in \cite{RosalesGuamn2023}. 
%
%
Regarding the characteristics of IICs, it is worth mentioning the smaller number of students per course, whose enrollment is voluntary and exclusively dedicated. These conditions could influence the motivation and commitment of students to their learning, see \cite{Scott2003}. Another notable characteristic is the role of the teacher during the development of IICs, which should not be limited to imparting content, but also include being a facilitator who encourages participation through active learning \cite{Scott2003}. In addition, the focus group mentioned that, in terms of assessment in the IICs, it is necessary to constantly monitor learning, first through a diagnostic assessment that evaluates the student's prior knowledge before beginning the IICs, and then by integrating formative assessments in a continuous and systematic manner, allowing the teacher to make immediate adjustments in their teaching-learning process, as mentioned in \cite{redish2006reverse}, also considering summative assessments during the process, on a weekly basis, which allow for the evaluation of student performance, that is, quantifying the degree to which the learning outcomes of the subject have been achieved. 
%

Finally, the importance of formative assessment as a highly effective tool for monitoring and tracking learning of the three types of educational content (being, doing, and knowing) is highlighted, as described in \cite{kohlmyer2009tale}. These assessment strategies are consistent with the principles of rapid and continuous feedback necessary in an IIC, bearing in mind that its intensive nature requires both students and teachers to maintain a constant pace of work, where timely feedback is a key factor in ensuring understanding of the content in a short period of time, which is in accordance with what is described in \cite{Daniel2000ARO, Scott2003}.

Regarding the quantitative results of the quasi-experiment, it can be observed that both groups, EG and CG, had a similar percentage of achievement in the BEMA pre-test at the beginning of the course, showing that they started the quasi-experiment under similar conditions (see Figure 4). Then, when comparing the results of the BEMA pre-test and post-test for each group, the EG achieved a significant improvement in its final performance, while the CG, although it achieved an improvement, this change was smaller (see Figure 7). When comparing the performance of both groups at the end of the quasi-experiment (see Figure 6), differences can be observed, but they are not significant. This result is expected over time, given that the CG was also in a teaching-learning process.


Another aspect to note is that the Mean of Individual Gains($<\text{g}>$) for the BEMA test was better for the EG compared to the CG (see Figure \ref{tabla2} and Figure \ref{tabla3}), both at the overall test level and for Hake's Gain (g) for each of the thematic units covered by the test (see Figure \ref{grafico3}), the latter showing that there is an improvement in the performance of EG students in each thematic unit of the BEMA test, unlike the CG, which only shows better performance in four of the eight thematic units of the test (see Figures \ref{grafico1} and \ref{grafico2}). It can be observed (see Figure \ref{grafico3}) that in the CG, in four of all thematic units (electric current and resistance, electrical potential, direct-current circuits and Gauss's law), showed a negative Hake´s Gain (g), i.e., with worse results in the post-test than in the pre-test. This could be due to many factors, one of which is random responses by some students, as analyzed by Alconchel et al. \cite{alconchel2015cuestionarios}. Although this analysis is outside the scope of the research questions, in order to avoid random responses in the BEMA, the post-test assessments were linked to a percentage of the final summative test of the course for both groups.

In terms of comparing the BEMA test scores in this study with other published research, this cannot be exhaustively done, since at the time of completion of this study, no research was found in the literature referring to the use of this test in remedial IICs. However, Alconchel et al. \cite{alconchel2015cuestionarios} administered the BEMA test at a Spanish university during a regular semester, obtaining scores higher than those recorded in the CG and EG of this study. One of the factors that could explain these differences is the fact that Chilean university students perform less well in OECD tests such as PISA 2022 \cite{OECD2023} and previous ones in the areas of language, mathematics and natural sciences, showing these characteristics at the time of entering university. 

%
Other elements to consider could be, first, the remedial nature of these courses, where enrollment is made up of students who have previously failed the subject in the regular semester; second, the fact that the IICs are given at a faster pace than a regular course. Finally, it should be noted that there are students who may not yet have managed to assimilate the pace of university life to learn effectively in their subjects. These aspects related to academic performance have been addressed in the literature, considering their different dimensions in Chilean universities \cite{vazquezanalisis, amestica2021efectos}.

Given the results obtained during the research process, especially the quantitative results of stage 3, it is possible to affirm that the innovative methodological proposal in the EG, through the use of blended learning methodologies that integrate interactive lectures with active learning activities and modern assessment strategies, allows for more efficient learning in an IIC on electromagnetism.

\section{CONCLUSIONS}

Although various studies on physics teaching have demonstrated the effectiveness of active methods in introductory STEM courses, most of these studies have been conducted exclusively in the context of regular courses, while scientific evidence for IICs is limited. To address this gap in the literature, this study explored the formulation, implementation, and efficiency of an innovative pedagogical proposal for an Electromagnetism IIC. The results obtained show how the combination of various active methodologies, together with formative assessment, contributes to significantly greater efficiency in conceptual understanding of electromagnetism compared to classes that did not integrate these strategies, as has been amply demonstrated in regular courses.

In relation to the results obtained in the focus group, it is concluded that the methodologies, types of assessment, and teacher-student interaction are consistent with the approach to teaching in regular courses in the current university context. It should also be noted that this information-gathering technique is a powerful qualitative research tool that is extremely useful for designing innovative proposals and enables evidence-based decision-making.

As for the conclusions of the quasi-experiment, the EG obtained better results than the CG in the Didactic Efficiency indicator (Hake's Gain) for the course, when compared with itself both globally and in the different thematic units of the BEMA test. However, no significant differences were found between the groups, which shows that both groups were immersed in a teaching-learning process.

Two original contributions of this work have been, first, the application of the BEMA conceptual test in an IIC on Electromagnetism and, second, the calculation of Hake's Gain to measure Didactic Efficiency in learning based on BEMA results for an IIC. In addition, it was confirmed that Hake's Gain is a useful indicator for quantifying the level of learning achieved in an IIC, as well as for comparing results between students at both the individual and group levels. These findings invite future research to define categorization ranges for Hake's Gain using the BEMA test since, for example, there are currently defined categories using the FCI that have been established in previous studies by North American and European universities. Given that the BEMA test is also conceptually more complex than the FCI, it is necessary to establish a specific categorization for Hake's Gain using its scores for both regular courses and IICs.

In general, to move from traditional pedagogical approaches to active learning practices, it is necessary to overcome considerable resistance on the part of teachers to the use of these methodologies. To facilitate this transition, it is essential to promote and highlight the evidence supporting the effectiveness of active methodologies in learning. Therefore, the authors suggest continuing with similar studies that integrate the realities of each institution, identifying the most effective methods, techniques, and strategies for learning in IICs, considering the current context of university teaching. It is also suggested to continue training higher education teachers through courses or workshops on various active methodologies for teaching physics, creating repositories available to teachers of both activities that use active methodological strategies and key questions for formative assessment.

It should also be noted that this research does not address the influence of other factors associated with the teaching and learning process, such as the role of the teacher during the learningbema process, motivation to learn, and gender, which may be variables to be examined in future studies to enrich the findings and propose new lines of intervention. It should also be mentioned that one of the advantages of PER in the context of IICs is that they are more motivating for research, as they are shorter in duration and results can be observed in less time.


In summary, intervention based on active methodological strategies and modern assessment has proven to be effective in an intensive teaching context, although there are areas for improvement that could be explored in future research or considered for other disciplines as well. The results of this study show that by making adjustments to different elements of the classroom, it is possible to improve teaching efficiency in IICs, demonstrating that it is very important to constantly review teaching practices according to the context, advances in education, and one's own reflection.

\section{ACKNOWLEDGEMENTS}

This study was conducted within the framework of the 2023 Teaching Project Development Fund (FDPD 2023) project, with the support of the Teaching Innovation in Engineering Unit (UIDIN) at the Catholic University of the North (UCN). Special thanks to Professor Claudio Ramírez Michea, MSc, for his collaboration in the success of this research. The authors also would like to thank all the teachers who participated in the Focus Group and the students enrolled in the 2024 Intensive Summer Course.
The author, M. Vallejo, dedicates this work to the memory of her son, Pablo Valenzuela, who passed away as this research was reaching its conclusion and whose light continues to accompany this journey.
 
\bibliography{referencesURL}

\section{APPENDIX} \label{apéndice}





%



%

\subsection{Activities in the classroom}


This appendix shows the log of activities carried out during the three weeks of the intensive course on Electromagnetism course on which this research was based. Three figures are presented describing the strategies and methodologies carried out each day during the, shown by week.

\begin{figure*}
\begin{center}
\begin{tabular}{ c }
 \includegraphics[width=0.8
\textwidth]{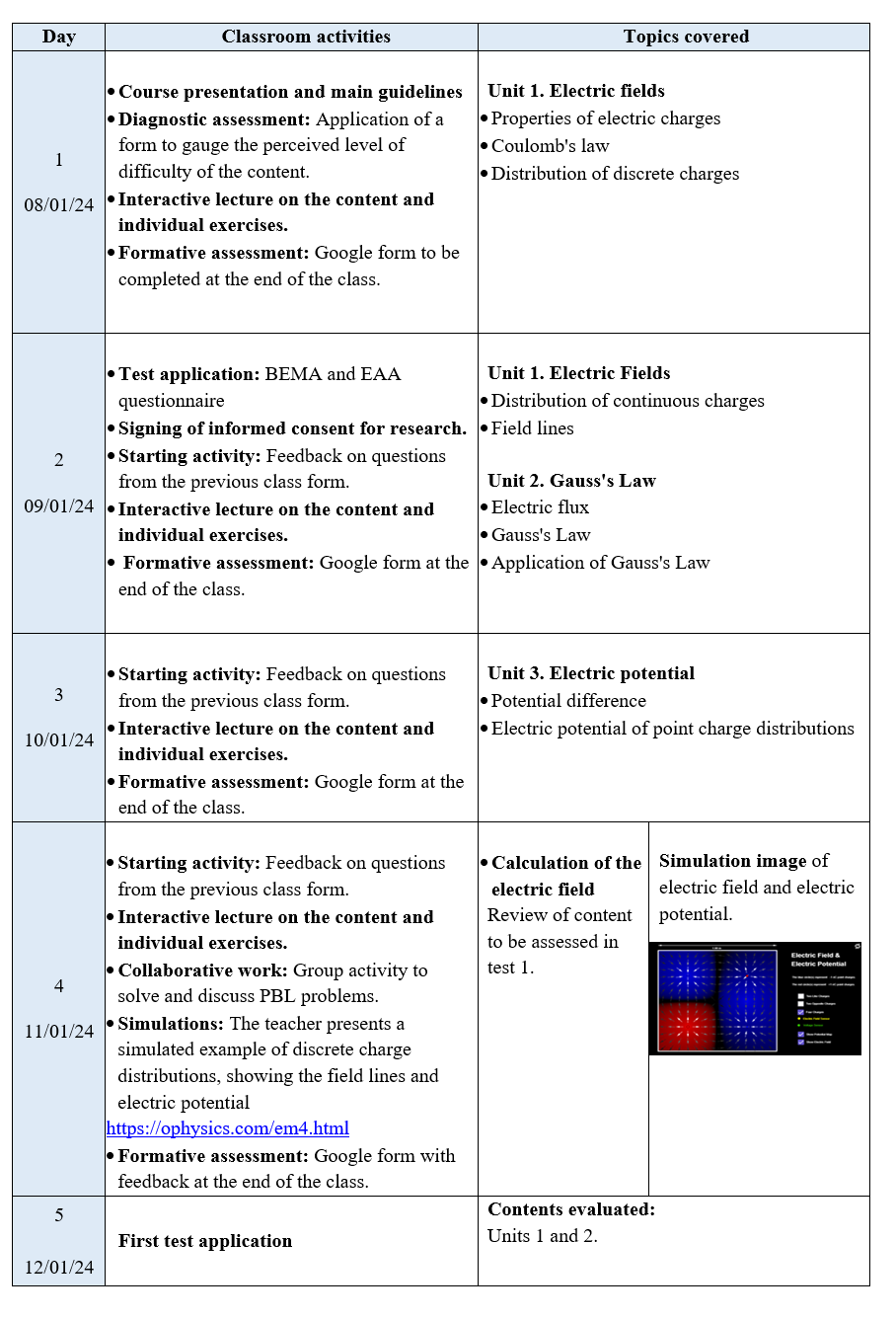} \\     
\end{tabular}
\end{center}
     \caption{Log of activities carried out during the intensive course on Electromagnetism: Week 1}
    \label{bitácora_semana1}
\end{figure*}

 \begin{figure*}
\begin{center}
\begin{tabular}{ c }
 \includegraphics[width=0.8
\textwidth]{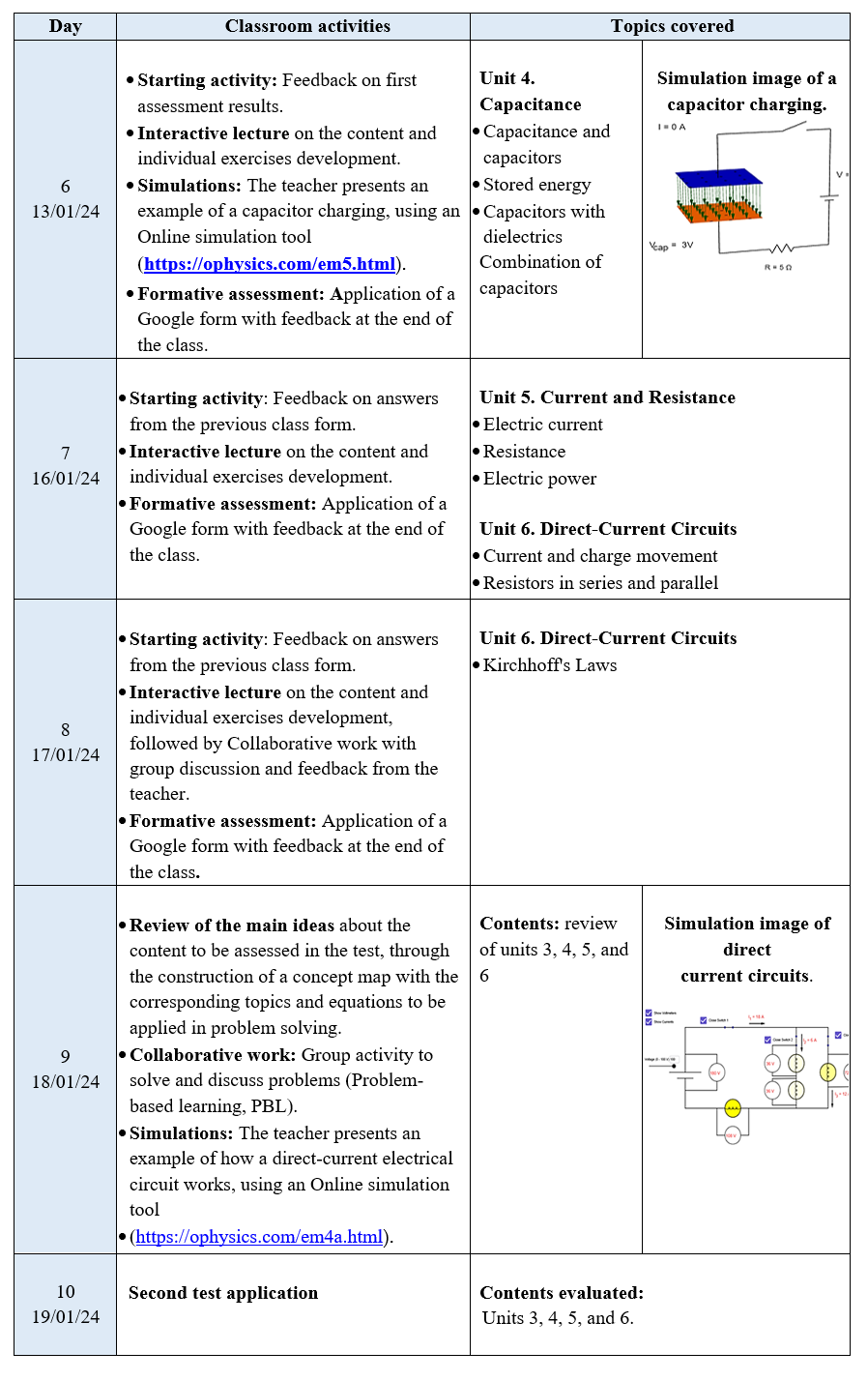} \\     
\end{tabular}
\end{center}
     \caption{Log of activities carried out during the intensive course on Electromagnetism: Week 2}
    \label{bitácora_semana2}
\end{figure*}

 \begin{figure*}
\begin{center}
\begin{tabular}{ c }
 \includegraphics[width=0.77
\textwidth]{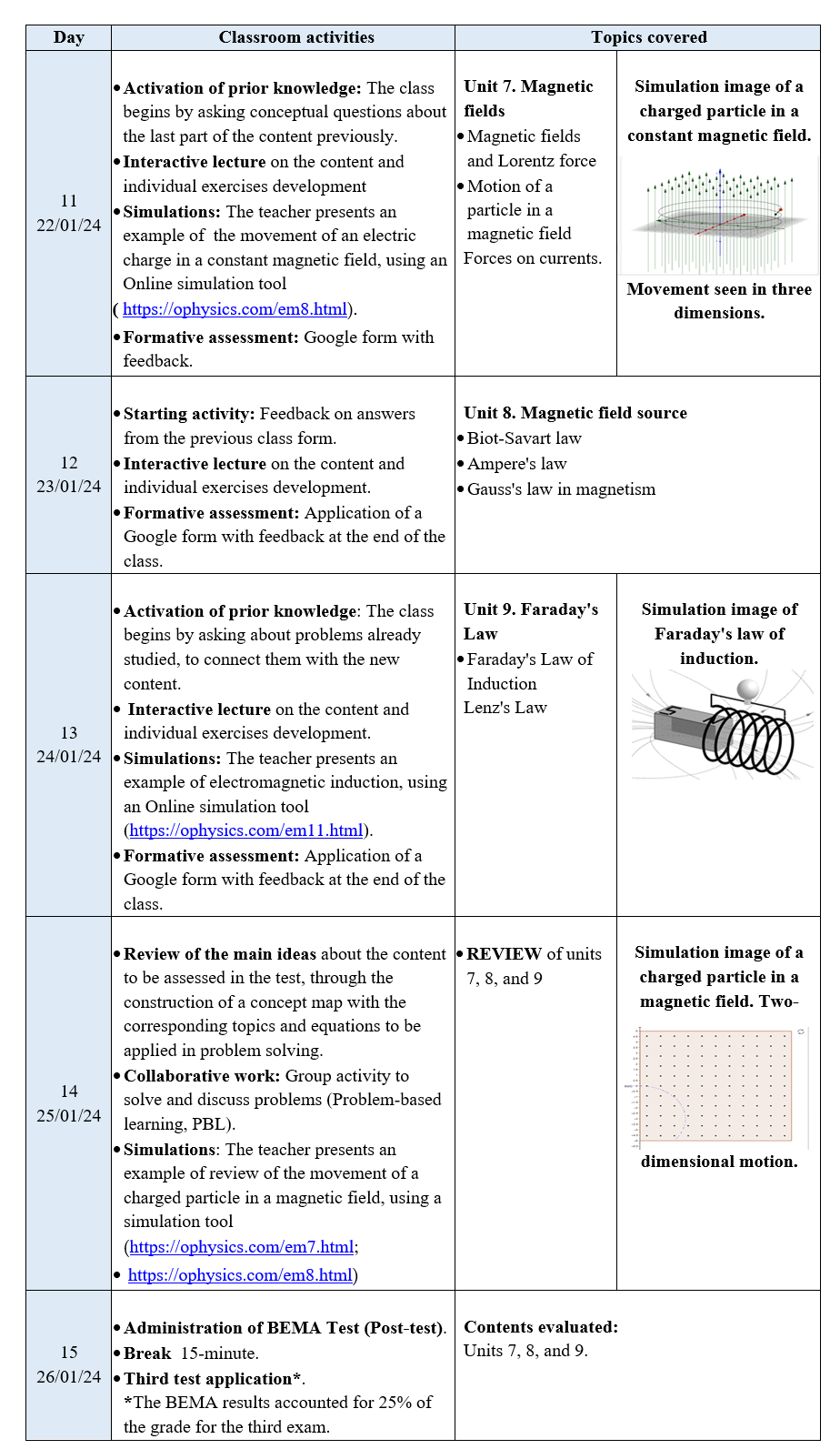} \\     
\end{tabular}
\end{center}
     \caption{Log of activities carried out during the intensive course on Electromagnetism: Week 3}
    \label{bitácora_semana3}
     \end{figure*}
\end{document}